\documentclass[a4paper,12pt]{article}

\pdfoutput=1

\usepackage[hmargin=.71in,vmargin=1.1in]{geometry}
\usepackage{indentfirst}
\usepackage{amsfonts}
\usepackage{amssymb}
\usepackage{mathrsfs}
\usepackage{upgreek}
\usepackage{amsmath}
\usepackage{mathtools}
\usepackage{mathabx}
\usepackage{tensor}
\usepackage{physics}
\usepackage{braket}
\usepackage{slashed}
\usepackage{authblk}
\usepackage{cite}
\usepackage{xcolor}
\usepackage{bm}
\usepackage{color,soul}
\usepackage{csquotes}
\usepackage{caption}
\usepackage{subcaption}
\usepackage{multirow}
\usepackage{graphicx}

\usepackage[bookmarksnumbered=true,bookmarksopen=true]{hyperref}
\hypersetup{colorlinks,
             linkcolor=[rgb]{0,0.3,0.6}, 
             citecolor=[rgb]{0,0.3,0.6}, 
             urlcolor=[rgb]{0,0.3,0.6}}

\newcommand\me{\mathrm{e}}
\newcommand{\dif}{\mathrm{d}}

\usepackage{booktabs}
\usepackage{multirow}

\usepackage{tikz}
\usetikzlibrary{matrix}

\begin{document}

\title{\Large\textbf{The impact of plunging matter on black-hole waveform}}

\author[a]{Ying-Lei Tian\thanks{yl.tian@mail.nankai.edu.cn}}
\author[b]{Hao Yang\thanks{hyang@ucas.ac.cn}}
\author[c]{Chen Lan\thanks{stlanchen@126.com}}
\author[a]{Yan-Gang Miao\thanks{Corresponding author: miaoyg@nankai.edu.cn}}

\affil[a]{\normalsize{\em School of Physics, Nankai University, 94 Weijin Road, Tianjin 300071, China}} 
\affil[b]{\normalsize{\em School of Fundamental Physics and Mathematical Sciences, Hangzhou Institute for Advanced Study, University of Chinese Academy of Sciences, Hangzhou 310024, China}}
\affil[c]{\normalsize{\em Department of Physics, Yantai University, 30 Qingquan Road, Yantai 264005, China}}

\date{ }

\maketitle

%%%%%%%%%%%%%%%%%%%%%%%%%%%%%%%%%%%%%%%%%%%%%%%%%%%%%%%%%%%%
\begin{abstract}
In this work, we introduce a novel framework to investigate ringdown gravitational waveforms in the presence of dynamical matter fields outside the horizon of a black hole. 
We systematically analyze two distinct scenarios of dynamical matter fields: motion along geodesics and uniform motion with constant velocity. 
Our results reveal rich phenomenology in the ringdown gravitational wave signals, including the suppression or enhancement of echoes, frequency shifts in the decay oscillations, and intricate modulations of the power-law tails.
Notably, we demonstrate that subluminal moving potentials can produce irregular echo patterns and shift the dominant frequencies, offering potential new observational signatures beyond the already-known ringdown analyses.
This study provides a new perspective for probing dynamic environments around black holes and offers a theoretical foundation for interpreting possible deviations in future gravitational wave detections.
\end{abstract}

%%%%%%%%%%%%%%%%%%%%%%%%%%%%%%%%%%%%%%%%%%%%%%%%%%%%%%%%%%%%
\tableofcontents

%%%%%%%%%%%%%%%%%%%%%%%%%%%%%%%%%%%%%%%%%%%%%%%%%%%%%%%%%%%%
\section{Introduction}
%%%%%%%%%%%%%%%%%%%%%%%%%%%%%%%%%%%%%%%%%%%%%%%%%%%%%%%%%%%%

Since the landmark detection of the GW150914 event by LIGO in 2015 \cite{LIGOScientific:2016aoc}, gravitational wave astronomy has emerged as a transformative window into strong-field gravity and extreme physics. 
Current research focuses predominantly on compact binary coalescence, which characteristically evolves through three distinct phases: inspiral, merger, and ringdown. 
While the inspiral and merger phases require large-scale numerical relativity simulations, the ringdown phase, owing to its linear perturbation nature, is particularly amenable to description through black hole perturbation theory \cite{Berti:2009kk,Konoplya:2011qq}. 
Within this framework, gravitational perturbations obey a Schr\"odinger-like wave equation with a specific effective potential. 

In static and spherically symmetric vacuum spacetimes, gravitational perturbations decompose into two fundamental modes: axial and polar. 
The axial perturbations are governed by the Regge-Wheeler equation \cite{Regge:1957td,Chandrasekhar:1985kt}, while polar perturbations are described by the Zerilli equation \cite{Zerilli:1970se,Zerilli:1970wzz}. 
The effective potentials of these equations not only determine the quasinormal mode (QNM) spectra but also encode profound information about spacetime. Recent studies have explored novel physical effects by modifying these effective potentials (a phenomenological approach that deviates from exact vacuum solutions yet remains physically insightful), such as introducing potential discontinuities \cite{Qian:2020cnz,Liu:2021aqh} or additional potential terms\cite{Chandrasekhar:1975zza,Cardoso:2019mes,Leung:1999iq}.  
This methodology proves valuable for investigating both environmental matter effects around black holes \cite{Barausse:2014tra,Cheung:2021bol,Berti:2022xfj,Guo:2022umh} and quantum-gravity-induced horizon microstructure \cite{Cardoso:2019apo,Li:2019kwa,Chakravarti:2021clm}.

More rigorously, the physical effects of environmental matter are modeled from the first principle through mass distributions or stress-energy tensors. 
This approach becomes particularly crucial when studying environmental effects like dark matter halos \cite{Konoplya:2018yrp,Cardoso:2021wlq,Figueiredo:2023gas,Cardoso:2022whc,Speeney:2024mas,Spieksma:2024voy,Maeda:2024tsg,Pezzella:2024tkf,Lan:2025brn}. However, postulating specific mass distributions often leads to formidable mathematical complexity in characterizing key spacetime features, including horizon structure, photon sphere, innermost stable circular orbit (ISCO), and so on\cite{Maeda:2024tsg}. 
%Therefore, this work adopts \red{a phenomenological Model: a simplified approach to directly modify the effective potential}, which offers computational advantages while maintaining physical interpretability through equivalent matter field descriptions \cite{Cardoso:2019mqo,Chung:2021roh}.
Therefore, we adopt a phenomenological model in this work: A simplified framework that directly modifies the effective potential. This approach provides computational advantages while maintaining physical interpretability through equivalent matter field descriptions \cite{Cardoso:2019mqo,Chung:2021roh}.
%Therefore, this work adopts the simplified approach of directly modifying the effective potential, which offers computational advantages while maintaining physical interpretability through equivalent matter field descriptions \cite{Cardoso:2019mqo,Chung:2021roh}.

We particularly focus on axial gravitational perturbations. The Regge-Wheeler potential features a prominent barrier near the photon sphere radius, where gravitational perturbations are primarily excited \cite{Cardoso:2016rao,Cardoso:2016oxy,Cardoso:2017njb}. Introducing a stationary additional Gaussian bump at an appropriate position creates a double-barrier structure, causing multiple reflections of perturbation signals that ultimately generate characteristic black hole ``echoes'' in the waveform \cite{Cardoso:2016rao,Cardoso:2016oxy,Cardoso:2017cqb}. 
In fact, such echo phenomena may arise in any compact object with potential cavity structures\cite{Cardoso:2019mes,Conklin:2017lwb}, including horizonless exotic compact objects (ECOs) like wormholes \cite{Cardoso:2016rao,Bueno:2017hyj,Yang:2024prm}, gravastars \cite{Mazur:2001fv,Visser:2003ge}, boson stars \cite{Schunck:2003kk,Chung:2021roh}, fuzzballs \cite{Lunin:2002qf,Mathur:2012jk}, and firewalls \cite{Mathur:2012jk,Almheiri:2012rt}.

Considering dynamical bumps to model matter dynamics reveals new phenomena owing to the chasing effect between perturbations and moving barriers \cite{Wang:2018mlp,Chen:2019hfg}, with outcomes depending on the barrier's initial position and motion. The ISCO marks the smallest stable orbital radius for massive particles in spherically symmetric spacetimes \cite{Maeda:2024tsg}. 
Objects within the ISCO spiral inward rapidly due to gravitational radiation or perturbations, ultimately plunging into black holes. 
Our results show that the matter inside the ISCO inevitably falls into black holes regardless of initial velocity, with near-horizon dynamic matter moving fast enough to prevent echoes as perturbations cannot catch up. 
Conversely, the matter beyond the ISCO can maintain stable circular orbits without initial velocity, but given sufficient initial radial velocity, it will also plunge inward. 
The resulting potential cavity first contracts and then expands, producing distinctive waveform features like reduced echo numbers, shortened echo periods, and frequency shifts compared to stationary potential cases.

The paper is organized as follows: Section \ref{sec:stationary} introduces vacuum cases with a stationary additional bump; Section \ref{sec:dynamcs} systematically investigates moving bump models, including both geodesic and uniform constant-velocity motion, where the latter potentially arises from electromagnetic fields or accretion-driven processes; Section 
\ref{sec:concl} summarizes our main findings and discusses their implications for gravitational wave observations and near-horizon physics.

%%%%%%%%%%%%%%%%%%%%%%%%%%%%%%%%%%%%%%%%%%%%%%%%%%%%%%%%%%%%
\section{Gravitational waveform with a fixed bump}
\label{sec:stationary}
%%%%%%%%%%%%%%%%%%%%%%%%%%%%%%%%%%%%%%%%%%%%%%%%%%%%%%%%%%%%

In this section, we take the vacuum Schwarzschild black hole as the spacetime background, whose line element is given by
\begin{equation}
\label{eq:metric}
\dif s^2 = -f(r) \dif t^2 + f^{-1}(r) \dif r^2 + r^2 \dif \Omega^2,
\end{equation}
where the metric function is given by $f(r) = 1 - \frac{2M}{r}$ and $M$ is the mass parameter. 
The event horizon, photon sphere, and innermost stable circular orbit (ISCO) are located at $r_\mathrm{H} = 2M$, $r_\mathrm{ps} = 3M$, and $r_\mathrm{ISCO} = 6M$, respectively.
For a vacuum Schwarzschild black hole, the axial gravitational field perturbation can be used to describe the tiny rotation of the gravitational field in spacetime, and the corresponding wave function $\Phi$ satisfies a Schr\"odinger-like equation of the form \cite{Berti:2009kk,Konoplya:2011qq}
\begin{equation}
\frac{\partial^2 \Phi}{\partial x^2} - \frac{\partial^2 \Phi}{\partial t^2} - V_{\mathrm{RW}} \Phi = 0,
\end{equation}
where $x$ denotes the tortoise coordinate defined by $x(r) = r + 2M \ln\left(\frac{r}{2M} - 1\right)$, in which the event horizon is located at negative infinity, while the photon sphere and ISCO radius are given by $x_\mathrm{ps} = (3 - 2\ln2)M$ and $x_\mathrm{ISCO} = (6 + 2\ln2)M$, respectively.
The Regge–Wheeler potential \cite{Regge:1957td,Chandrasekhar:1985kt} is expressed as
\begin{equation}
V_{\mathrm{RW}}(r) = f(r) \left[ \frac{\ell(\ell+1)}{r^2} - \frac{6M}{r^3} \right].
\end{equation}

However, a vacuum represents an idealized condition. 
In reality, spacetime is invariably filled with various forms of matter.
Under such circumstances, the effective potential of the axial gravitational field perturbation is influenced by the matter in spacetime.
Therefore, on the basis of the Regge-Wheeler potential, we introduce an additional local time-independent Gaussian potential, 
\begin{equation}
V_{\mathrm{add}}(x(r)) = \frac{\epsilon}{M^2} \,\me^{-\frac{(x(r) - a)^2}{\sigma^2}},
\end{equation}
to characterize the influence of matter, which we also refer to as a bump, whose physical interpretation is provided in App.~\ref{app:physical_interp}. 
Here, $\epsilon$, $\sigma$, and $a$ denote the amplitude, width, and peak position of the Gaussian profile, respectively, where $\sigma$ and $a$ are defined in the tortoise coordinate. 
Then, the total effective potential governing the perturbation becomes
\begin{equation}
V_{\mathrm{eff}}(r) = V_{\mathrm{RW}}(r) + V_{\mathrm{add}}(x(r)).
\end{equation}
This additional term can be interpreted as an effective model for near-horizon quantum structures or external matter environments such as accretion disks or dark matter halos \cite{Cheung:2021bol}.

\begin{figure}[!ht]
	\centering
    \begin{subfigure}[b]{0.9\textwidth}
        \centering
        \includegraphics[width=\textwidth]{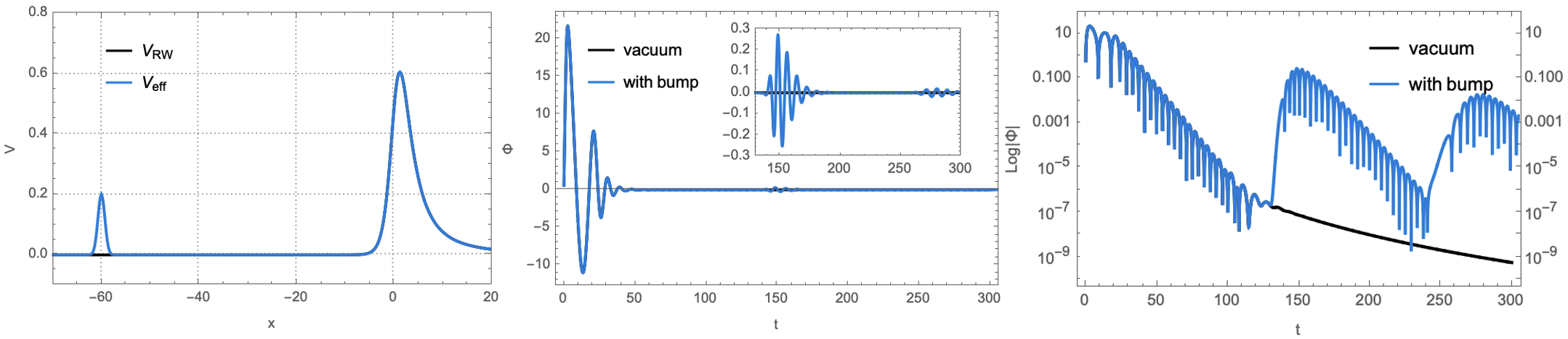}
        \caption{$a=-60$}
        \label{fig:fixed_n60}
    \end{subfigure}
        
    \begin{subfigure}[b]{0.9\textwidth}
        \centering
        \includegraphics[width=\textwidth]{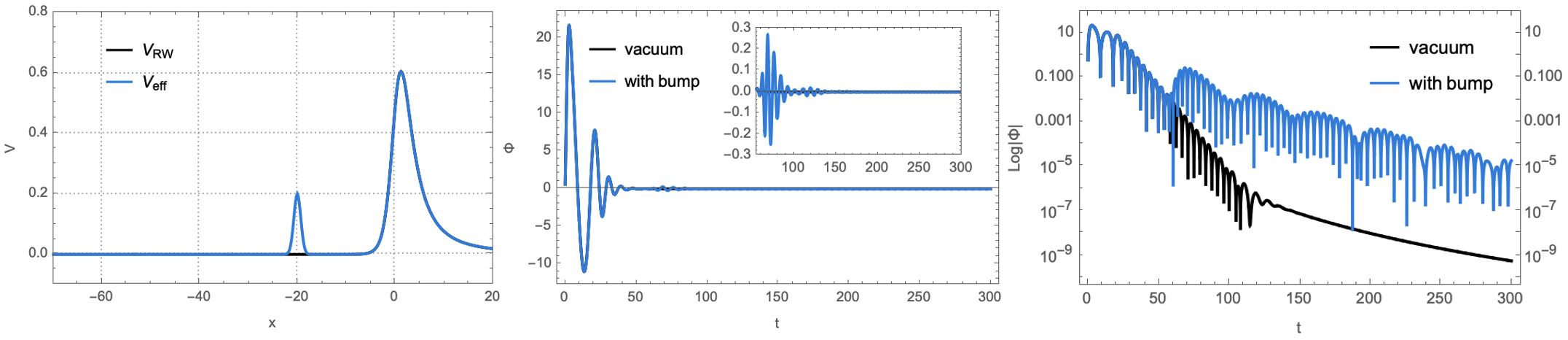}
        \caption{$a=-20$}
        \label{fig:fixed_n20}
    \end{subfigure}

    \begin{subfigure}[b]{0.9\textwidth}
        \centering
        \includegraphics[width=\textwidth]{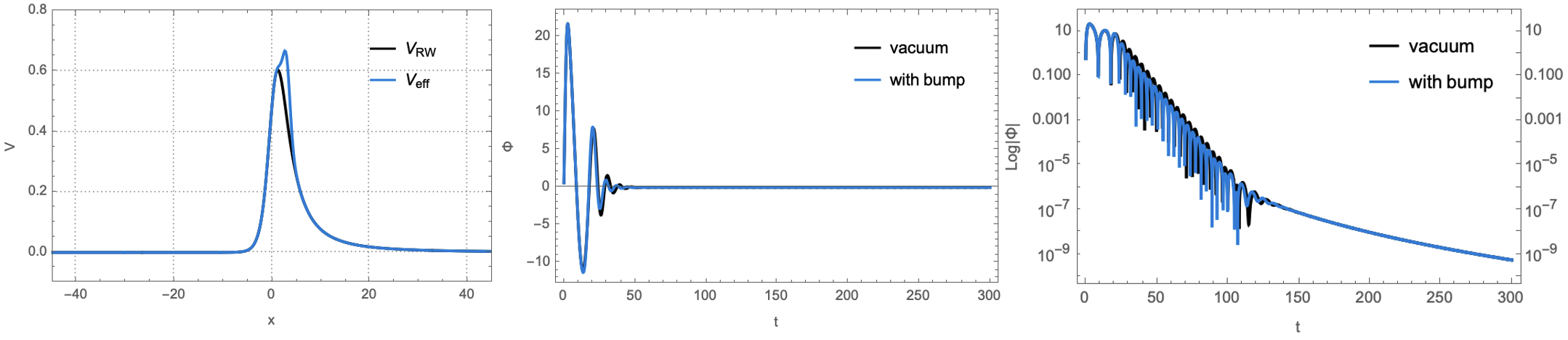}
        \caption{$a=3$}
        \label{fig:fixed_p3}
    \end{subfigure}
    
    \begin{subfigure}[b]{0.9\textwidth}
        \centering
        \includegraphics[width=\textwidth]{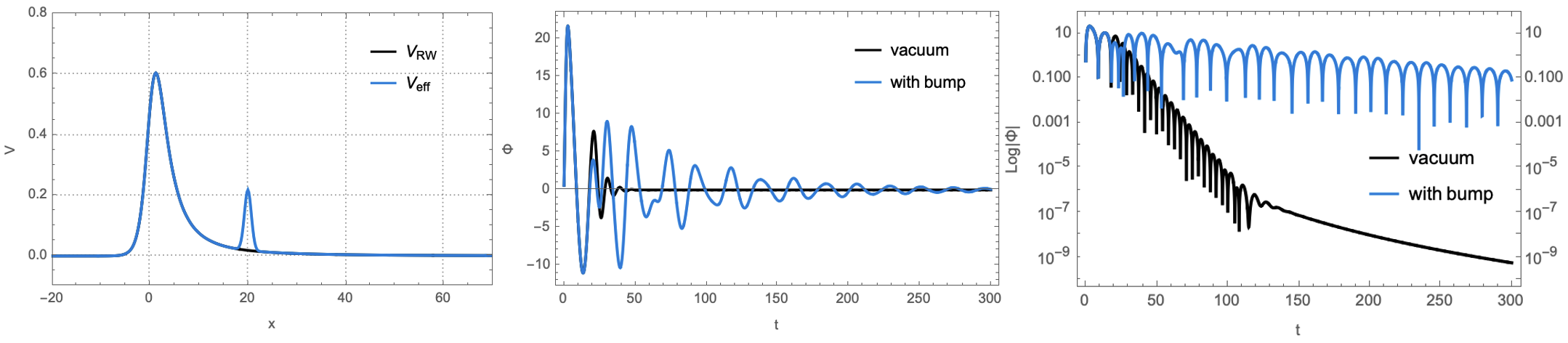}
        \caption{$a=20$}
        \label{fig:fixed_p20}
    \end{subfigure}

    \begin{subfigure}[b]{0.9\textwidth}
        \centering
        \includegraphics[width=\textwidth]{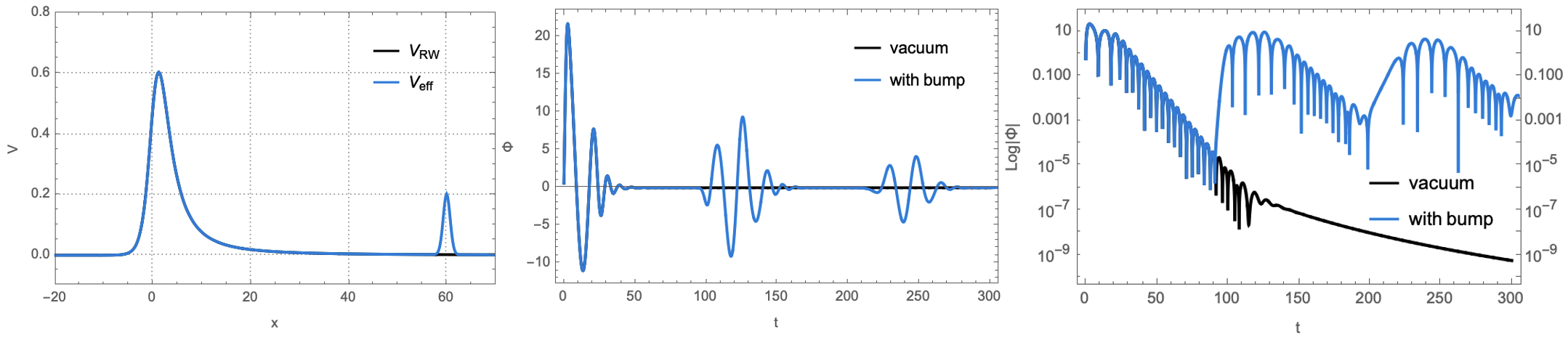}
        \caption{$a=60$}
        \label{fig:fixed_p60}
    \end{subfigure}

	\captionsetup{width=.9\textwidth}
	\caption{In the left column, we present the relative positions of the Gaussian bump and the Regge-Wheeler potential corresponding to different values of $a$. Meanwhile, we present the time-domain ringdown waveforms in the middle column and the corresponding logarithmic plots in the right column for various values of the parameter $a$. }
	\label{fig:fixed}
\end{figure}

In our analysis, we focus on several representative cases, which are distinguished by the different positions of the Gaussian bump relative to the Regge-Wheeler potential peak.
%For numerical consistency, we take the mass parameter as $2M=1$.
%\red{And consider the simple situation where $\ell=2$.}
%Then, we fix the amplitude and width of the Gaussian bump to \red{$\epsilon = 0.05$} and $\sigma = 1$, respectively, and explore the ringdown waveform for multiple values of $a \in \{-60, -20, 3, 20, 60\}$.
For numerical consistency, we set the mass parameter to $2M=1$ (therefore the photon sphere at $ 3/2 - \ln2$ and the ISCO at $3 + \ln2$) and focus on the simple case of $\ell=2$. The amplitude and width of the Gaussian bump are fixed at $\epsilon = 0.05$ and $\sigma = 1$, respectively. We then investigate the ringdown waveforms for several bump locations, $a \in {-60, -20, 3, 20, 60}$.
The values $a = -60, -20$ correspond to regions inside the photon sphere and near the horizon, which are used in \cite{Cardoso:2019apo,Li:2019kwa} to simulate quantum structures near the horizon. The value $a = 3$ lies between the photon sphere and ISCO, while $a = 20, 60$ correspond to regions outside the ISCO, far from the black hole, as employed in \cite{Liu:2021aqh,Barausse:2014tra,Cheung:2021bol} to model environmental matter around black holes.
For greater specificity, in the left column of Fig.~\ref{fig:fixed}, we present the relative positions of the Gaussian bump and the Regge-Wheeler potential corresponding to different values of $a$.
Meanwhile, in Fig.~\ref{fig:fixed} we present the time-domain ringdown waveforms in the middle column and the corresponding logarithmic plots in the right column for various values of the parameter $a$.

For $a \in \{-60, -20, 20, 60\}$, the introduction of an additional bump leads to the formation of a cavity-like structure between the peaks of the Regge–Wheeler potential and the bump. 
This cavity gives rise to echo signals in the resulting gravitational waveform.
%\red{The details of echo generation in the fixed bump model are presented in Table.~\ref{tab:fixed_bump_quantitative}, which demonstrate that echoes are produced with periodicity, exhibiting a time interval approximately twice the inter-peak separation, while their amplitudes gradually diminish.}
The details of echo generation in the fixed-bump model are summarized in Table~\ref{tab:fixed_bump_quantitative}. The results show that the echoes occur periodically, with time intervals roughly twice the inter-peak separation, and their amplitudes gradually decrease over time.
The echo signatures are relatively pronounced when the bump is located far from the peak of the Regge–Wheeler potential ($a=\pm 60$), indicating strong wave reflections between the two barriers, see Figs.\ \ref{fig:fixed_n60} and \ref{fig:fixed_p60}. 
These echoes manifest as secondary bursts of oscillations following the initial ringdown phase, particularly evident in the logarithmic plots. 
In contrast, when the bump is positioned closer to the main peak of the Regge–Wheeler potential, see Figs.\ \ref{fig:fixed_n20} and \ref{fig:fixed_p20}, the echo pattern becomes less distinct. 
And the corresponding waveforms display long-lived and slowly decaying oscillations, which are clearly visible in both the linear and logarithmic plots. 
These persistent oscillations indicate the presence of quasi-bound states trapped between the two barriers, a phenomenon reminiscent of quasi-stationary resonances.

\begin{table}[htbp]
\centering
\caption{Echoes for the fixed bump model.}
\label{tab:fixed_bump_quantitative}
\begin{tabular}{cccc}
\toprule
$a$ & echo number & time of appearance ($t/M$) & amplitude ratio (\%) \\
\midrule
\multirow{2}{*}{-60} & \multirow{2}{*}{2} & 130.8 & 1.602 \\
                     &                    & 242.1 & 0.115 \\
\midrule
\multirow{5}{*}{-20} & \multirow{5}{*}{5} & 68.65 & 1.581 \\
                     &                    & 119.7 & 0.115 \\
                     &                    & 162.5 & 0.017 \\
                     &                    & 205.2 & 0.00294 \\
                     &                    & 247.8 & 0.00052 \\
\midrule
3 & — & — & — \\
\midrule
20 & too ambiguous & — & — \\
\midrule
\multirow{2}{*}{60} & \multirow{2}{*}{2} & 91.0 & 56.46 \\
                    &                    & 199.2 & 26.60 \\
\bottomrule
\end{tabular}
\end{table}

For $a = 3$, the bump is positioned near the main barrier peak and leads to only minor modifications in the potential shape. Consequently, the waveform closely resembles the vacuum case, showing a typical ringdown followed by rapid decay without significant late-time oscillatory tails.

Another notable observation is the frequency asymmetry introduced by the position of the additional Gaussian bump. When the bump is placed to the right of the Regge–Wheeler potential peak (i.e., further from black holes), the dominant frequency of the waveform is significantly lower than that in the case where the bump is placed to the left (i.e., closer to the horizon). 
This suggests a sensitive dependence of the ringdown spectrum on the spatial configuration of the potential.

Finally, we analyze Fig.\ \ref{fig:fixed} from a different perspective. 
If we interpret subfigures \ref{fig:fixed_n60}, \ref{fig:fixed_n20}, \ref{fig:fixed_p3}, \ref{fig:fixed_p20}, and \ref{fig:fixed_p60} as a sequence of frames in an animation, where matter adiabatically moves towards a black hole due to its gravitational pull, the bump gradually shifts from the right to the left. 
Here, ``adiabatic" refers to the condition in which the Gaussian bump moves much more slowly than the characteristic velocity associated with the echo oscillations between the two potential barriers.
In this context, the waveform sequence can be viewed as the concatenation of subfigures \ref{fig:fixed_n60}, \ref{fig:fixed_n20}, \ref{fig:fixed_p3}, \ref{fig:fixed_p20}, and \ref{fig:fixed_p60}. 
This phenomenon naturally motivates our investigation into how a moving Gaussian potential modification influences the ringdown waveform.

%%%%%%%%%%%%%%%%%%%%%%%%%%%%%%%%%%%%%%%%%%%%%%%%%%%%%%%%%%%%
\section{Gravitational waveform with a dynamic bump}
\label{sec:dynamcs}
%%%%%%%%%%%%%%%%%%%%%%%%%%%%%%%%%%%%%%%%%%%%%%%%%%%%%%%%%%%%

We now extend our analysis to a time-dependent scenario in which the additional Gaussian bump dynamically evolves. 
To capture this physical picture, we introduce a moving bump modeled as
\begin{equation}
V_{\mathrm{add}}(x(r),t) = \frac{\epsilon}{M^2} \,\me^{-\frac{(x - a(t))^2}{\sigma^2}}.
\end{equation}
Here $a(t)$ describes the time-dependent position of the bump and includes two important parameters: $a_0=a(0)$ denoting the initial position.
%and $v$ the inward velocity. 
Therefore, the total effective potential thus becomes
\begin{equation}
V(r,t) = V_{\rm{RW}}(r) + V_{\rm{add}}(x(r), t).
\end{equation}
This setup effectively models matter or field configurations slowly accreting toward black holes.

%%%%%%%%%%%%%%%%%%%%%%%%%%%%%%%%%%%%%%%%%%%%%%%%%%%%%%%%%%%%
\subsection{Geodesic-inspired motion}
%%%%%%%%%%%%%%%%%%%%%%%%%%%%%%%%%%%%%%%%%%%%%%%%%%%%%%%%%%%%

We interpret the moving bump as being sourced by a matter field composed of particles following time-like geodesics. 
The four-velocity of particles is given by $u^\mu = \frac{\dif y^\mu}{d\tau}$, where $\tau$ is the proper time along the geodesic. In a spherically symmetric spacetime, there are two Killing vectors: $\xi^a = (\partial_t)^a$ and $\psi^a = (\partial_\varphi)^a$.
Therefore, the corresponding conserved energy and angular momentum of particles are $E = -\xi^a u_a$ and $L_z = \psi^a u_a$, respectively. 
The total angular momentum satisfies
\begin{equation}
L^2 = r^4 (u^\theta)^2 + r^4 \sin^2{\theta}\,(u^\varphi)^2.
\end{equation}
The explicit components of the four-velocity read
\begin{subequations}
\begin{equation}
u^t = \frac{\dif t}{\dif \tau} = \frac{E}{f}, \qquad u_t = -E,
\end{equation}
\begin{equation}
u^\varphi = \frac{\dif\varphi}{\dif\tau} = \frac{L_z}{r^2 \sin^2{\theta}}, \qquad u_\varphi = L_z,
\end{equation}
\begin{equation}
r^2 (u^\theta)^2 = L^2 - \frac{L_z^2}{\sin^2{\theta}}.
\end{equation}
\end{subequations}
Imposing the normalization condition: $u^\mu u_\mu = -1$, we obtain the radial component,
\begin{equation}
u^r = \frac{\dif r}{\dif\tau} = \sqrt{E^2 - f \left(1 + \frac{L^2}{r^2}\right)}.
\end{equation}
Expressed in terms of the tortoise coordinate $x$, the radial velocity relative to coordinate time $t$ reads
\begin{equation}
v(r) = \frac{\dif x}{\dif t} = \sqrt{1 - \frac{f}{E^2} \left(1 + \frac{L^2}{r^2}\right)}.
\end{equation}

\begin{figure}[!ht]
	\centering
	\begin{subfigure}[b]{0.45\textwidth}
		\centering
		\includegraphics[width=\textwidth]{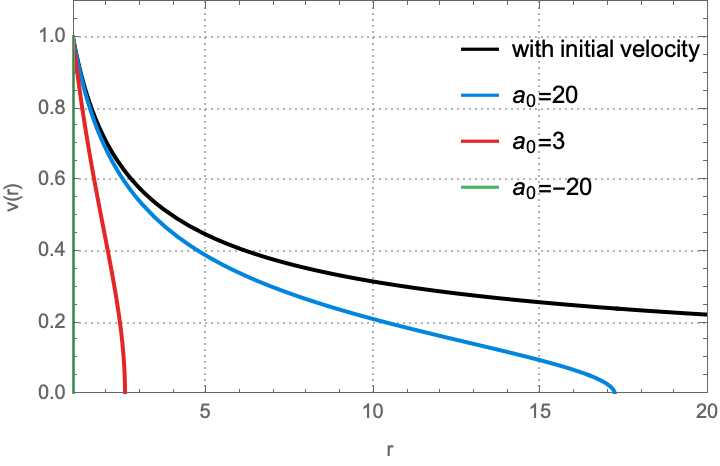}
		\caption{$v(r)$}
		\label{fig:V_radial}
	\end{subfigure}
	\begin{subfigure}[b]{0.45\textwidth}
		\centering
		\includegraphics[width=\textwidth]{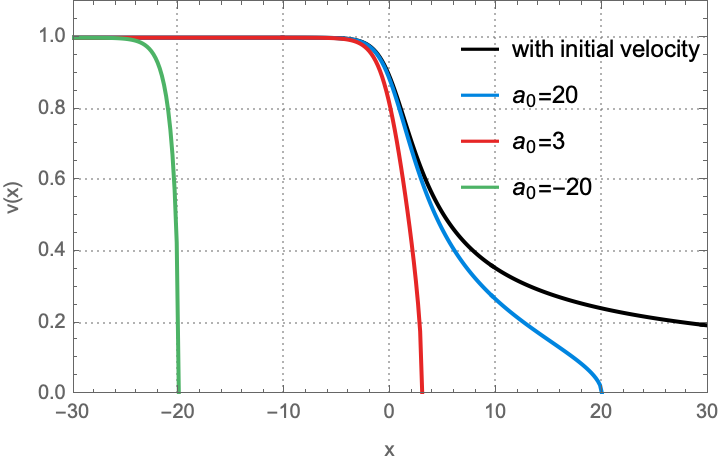}
		\caption{$v(x)$}
		\label{fig:V_tortoise}
	\end{subfigure}
	\captionsetup{width=.9\textwidth}
	\caption{The figure shows the velocity distributions referring to radial and tortoise coordinates. The black curves correspond to a free fall from infinity, while the blue, red, and green curves represent finite initial positions $a_0 = 20, 3, -20$, respectively. }
	\label{fig:V_distribution}
\end{figure}

For a purely radial infall ($L=0$), two scenarios arise:  
\begin{enumerate}
\item A free fall from infinity with an initial radial velocity:  
   $E = 1$, leading to
   \begin{equation}
   u^r = \sqrt{\frac{1}{r}}, \qquad v(r) = \sqrt{\frac{1}{r}}.
   \end{equation}
\item A free fall from a finite radius $r_0$ without initial radial velocity:  
   $E = \sqrt{1 - \frac{1}{r_0}}$, leading to
   \begin{equation}
   u^r = \sqrt{\frac{r_0 - r}{r_0 r}}, \qquad v(r) = \sqrt{\frac{r_0 - r}{(r_0 - 1)r}}.
   \end{equation}
\end{enumerate}
In Fig.~\ref{fig:V_distribution}, we present these velocity profiles in both radial ($r$) and tortoise ($x$) coordinates. 
In the near-horizon limit ($r \rightarrow 1$, $x \rightarrow -\infty$), all profiles asymptotically approach the speed of light ($v \rightarrow 1$), signifying that the bump eventually moves inward at nearly the luminal speed.

\begin{figure}[!ht]
	\centering
    \begin{subfigure}[b]{0.9\textwidth}
        \centering
        \includegraphics[width=\textwidth]{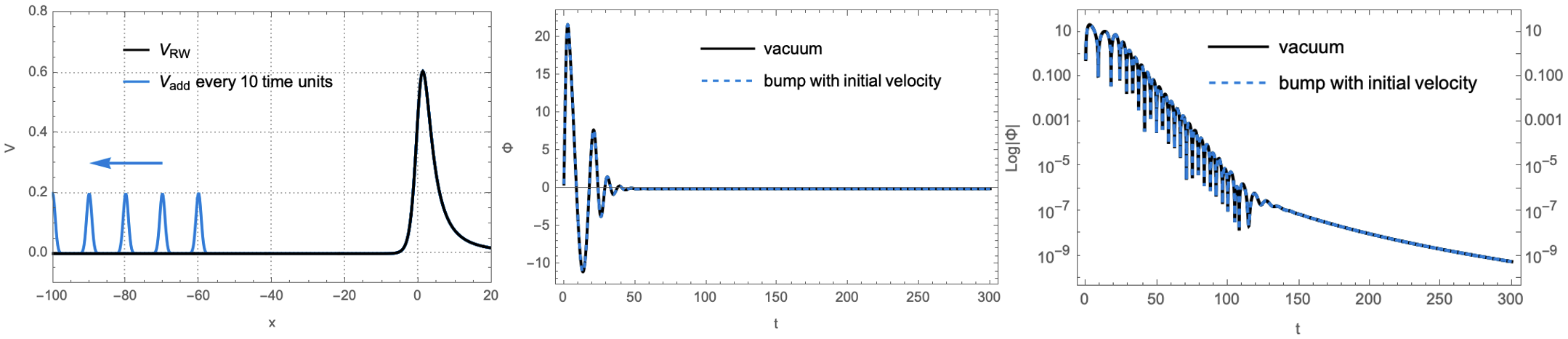}
        \caption{$a=-60$}
        \label{fig:Geodesic_n60}
    \end{subfigure}

    \begin{subfigure}[b]{0.9\textwidth}
        \centering
        \includegraphics[width=\textwidth]{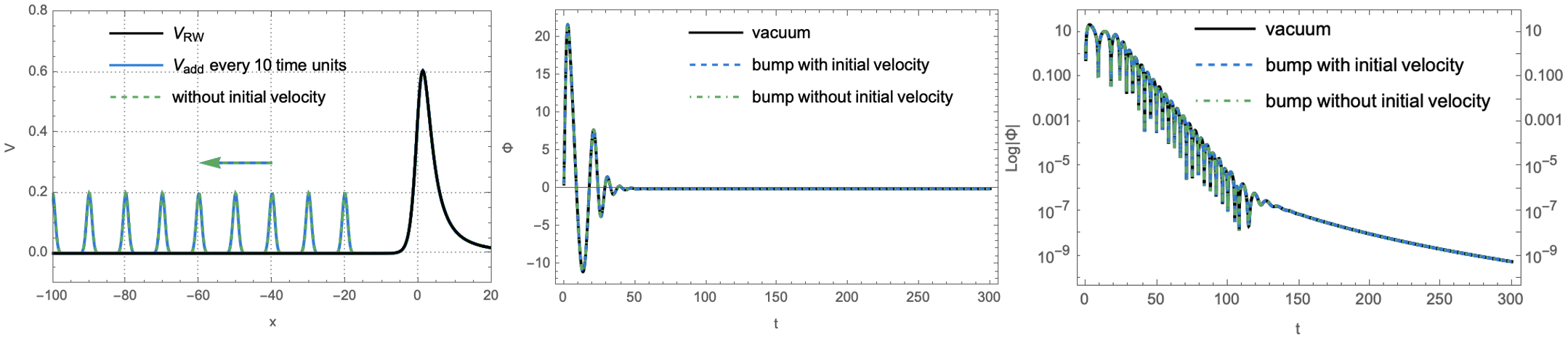}
        \caption{$a=-20$}
        \label{fig:Geodesic_n20}
    \end{subfigure}

    \begin{subfigure}[b]{0.9\textwidth}
        \centering
        \includegraphics[width=\textwidth]{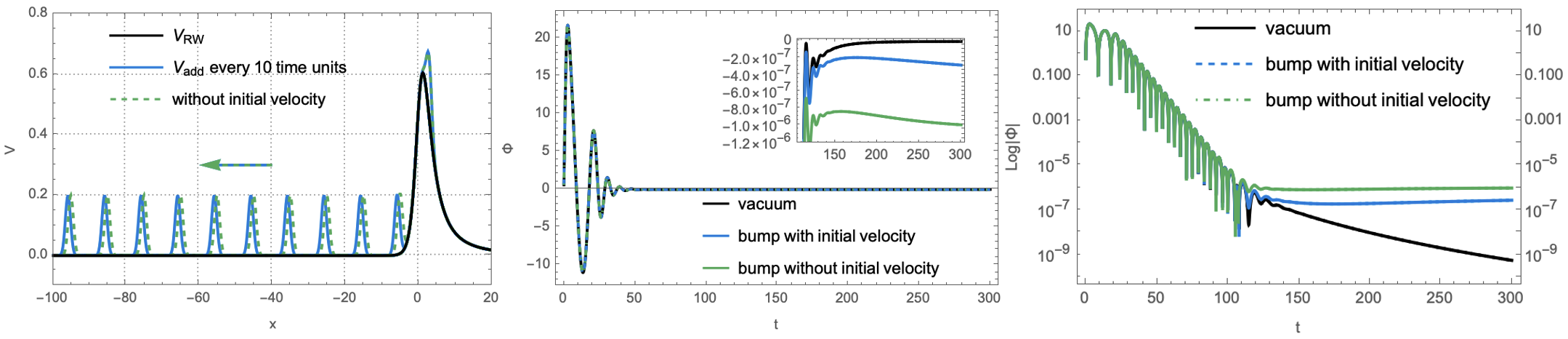}
        \caption{$a=3$}
        \label{fig:Geodesic_p3}
    \end{subfigure}

    \begin{subfigure}[b]{0.9\textwidth}
        \centering
        \includegraphics[width=\textwidth]{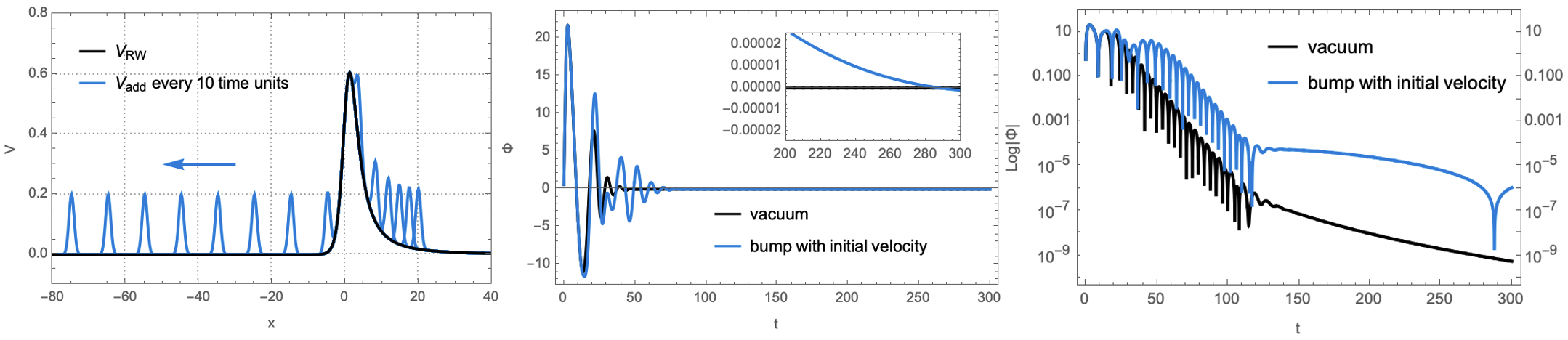}
        \caption{$a=20$}
        \label{fig:Geodesic_p20}
    \end{subfigure}

    \begin{subfigure}[b]{0.9\textwidth}
        \centering
        \includegraphics[width=\textwidth]{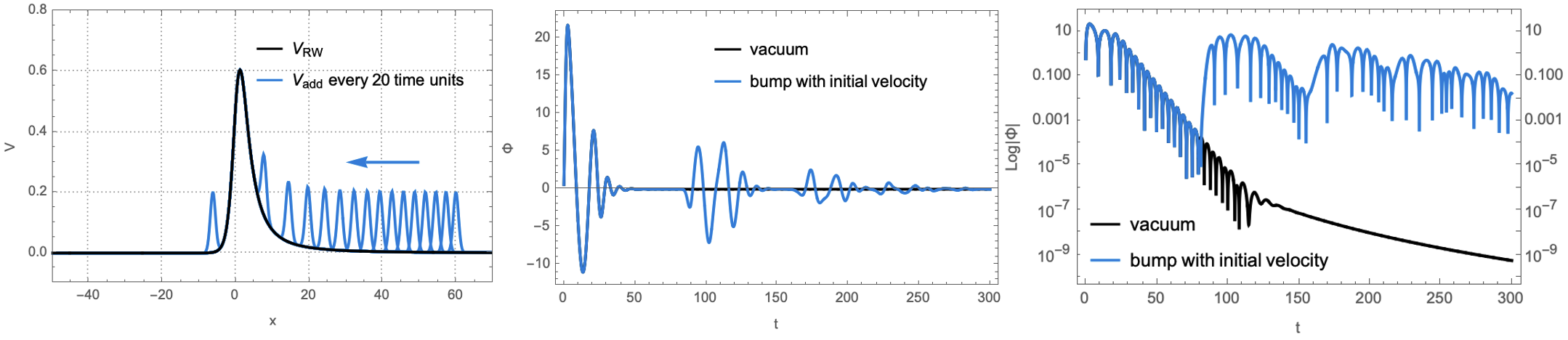}
        \caption{$a=60$}
        \label{fig:Geodesic_p60}
    \end{subfigure}
    
	\captionsetup{width=.9\textwidth}
	\caption{Waveforms of axial gravitational field perturbation with an additional bump moving along the geodesics. The blue and green curves correspond to the motion with and without the initial velocity, respectively. The black curve denotes the vacuum reference case.}
	\label{fig:moving_Geodesic}
\end{figure}

Next, we compute the axial gravitational perturbation waveforms in the presence of a dynamically moving bump following these geodesics. 
In Fig.~\ref{fig:moving_Geodesic}, we summarize our computed results, where the black curves correspond to the vacuum Schwarzschild case, the blue curves represent bumps with a free fall from infinity, and the green curves show bumps starting from a finite position $r_0=a_0$ with zero initial radial velocity. 
The first column in Fig.~\ref{fig:moving_Geodesic} illustrates the bump trajectories marked every 10 or 20 time units, and the arrows indicate the direction of motion. 
For $a_0 = -60$, in the case of a free fall from infinity, the velocity of the bump tends to the speed of light. 
For the case of a free fall starting from this point $r_0=a_0$, the velocity of the bump also tends toward the speed of light in a very short time. 
These two scenarios cannot be distinguished by their ringdown waveforms.
Therefore, we only present the scenario of a free fall from infinity in Fig.~\ref{fig:Geodesic_n60}.
For $a_0=20, 60$, the case of a free fall starting from this point $r_0=a_0$, the bumps do not show any significant movement during the time of our simulation, which leads to the waveforms being consistent with those in Figs.~\ref{fig:fixed_p20} and~\ref{fig:fixed_p60}.
Therefore, we also present only the scenario of a free fall from infinity in Figs.~\ref{fig:Geodesic_p20} and~\ref{fig:Geodesic_p60}.

\begin{table}[htbp]
\centering
\caption{Echoes for the geodesic motion model}
\label{tab:geodesic_bump_quantitative}
\begin{tabular}{cccccc}
\toprule
\multirow{2}{*}{$a$} & initial & echo & time of appearance & amplitude ratio & \multirow{2}{*}{supplement} \\
                     & velocity & number & ($t/M$) & (\%) &  \\
\midrule
-60 & nonzero & — & — & — & — \\
\midrule
\multirow{2}{*}{-20} & nonzero & — & — & — & — \\
                     & zero & — & — & — & — \\
\midrule
\multirow{2}{*}{3} & nonzero & — & — & — & slower decay in late-time tail \\
                   & zero & — & — & — & slower decay in late-time tail \\
\midrule
20 & nonzero & 1 & 30.6 & 25.42 & slower decay in late-time tail \\
\midrule
\multirow{3}{*}{60} & \multirow{3}{*}{nonzero} & \multirow{3}{*}{3+} & 81.4 & 40.55 & \multirow{3}{*}{—} \\
                    & & & 156.2 & 13.35 & \\
                    & & & 216.2 & 4.48 & \\
\bottomrule
\end{tabular}
\end{table}

%The second and third columns in Fig.~\ref{fig:moving_Geodesic} depict the corresponding waveforms\red{, and the echo generation details of the geodesic motion model are presented in Table.~\ref{tab:geodesic_bump_quantitative}, which demonstrate that echoes are either significantly reduced or completely absent}. 
The second and third columns of Fig.~\ref{fig:moving_Geodesic} display the corresponding waveforms, while Table~\ref{tab:geodesic_bump_quantitative} summarizes the quantitative results. It is evident that the echoes are either strongly suppressed or entirely absent in the geodesic motion model.
In the $a_0 = -60, -20$ cases (Figs.~\ref{fig:Geodesic_n60} and \ref{fig:Geodesic_n20}), echo signature vanishes as the bump moves inward nearly at light speed, preventing the perturbation signal from transmission and reflecting. 
For $a_0=3$ (Fig.~\ref{fig:Geodesic_p3}), the proximity of the bump to the main peak modulates the late-time power-law tail, leading to slower decay and amplitude differences.  
For $a_0 = 20, 60$ (Figs.~\ref{fig:Geodesic_p20} and \ref{fig:Geodesic_p60}), the bump initially creates a cavity, allowing for one or a few early echoes. 
However, as the bump moves toward the horizon, the cavity diminishes or disappears.
Therefore, compared to the stationary bump situation, the moving bump causes the echo signal to weaken and eventually disappear.
In particular, for $a_0=60$, we can observe that the intervals and amplitudes of the echo signals are gradually decreasing, eventually tending towards a damping oscillation mode.
Overall, these results demonstrate that the dynamics of a bump play a crucial role in shaping the waveform structure, especially in controlling the presence and characteristics of echo signals.

%%%%%%%%%%%%%%%%%%%%%%%%%%%%%%%%%%%%%%%%%%%%%%%%%%%%%%%%%%%%
\subsection{Constant velocity motion}
%%%%%%%%%%%%%%%%%%%%%%%%%%%%%%%%%%%%%%%%%%%%%%%%%%%%%%%%%%%%

In contrast to the geodesic-inspired motion discussed previously, we consider a simplified scenario here, where the additional Gaussian bump moves inward with a prescribed constant velocity $v < 1$.
Because a strictly constant subluminal velocity allows us to systematically investigate the impact of slower-moving bumps on the perturbation waveform, including the emergence of multiple echoes and modifications to the late-time decay oscillations.

\begin{figure}[!ht]
	\centering
    \begin{subfigure}[b]{0.9\textwidth}
        \centering
        \includegraphics[width=\textwidth]{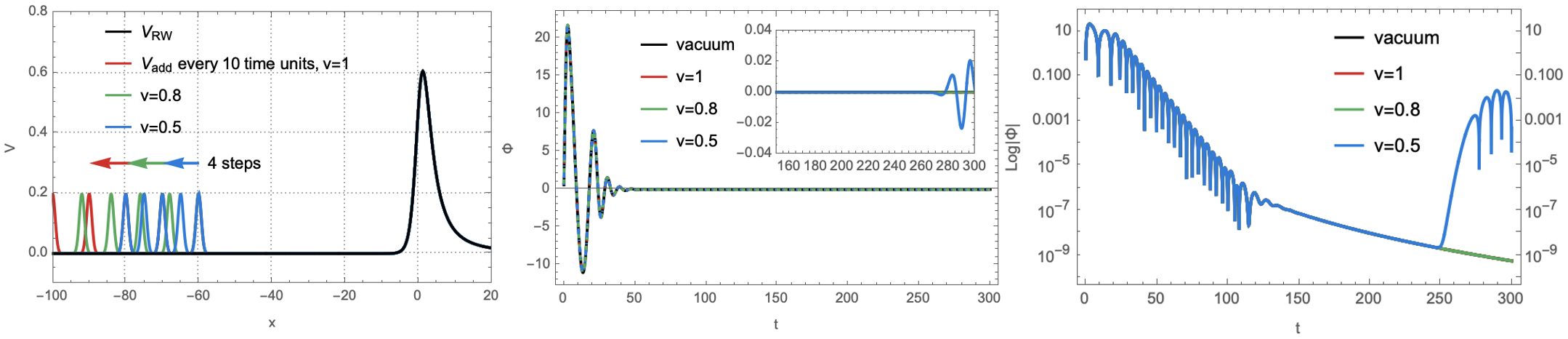}
        \caption{$a=-60$}
        \label{fig:ConstantV_n60}
    \end{subfigure}

    \begin{subfigure}[b]{0.9\textwidth}
        \centering
        \includegraphics[width=\textwidth]{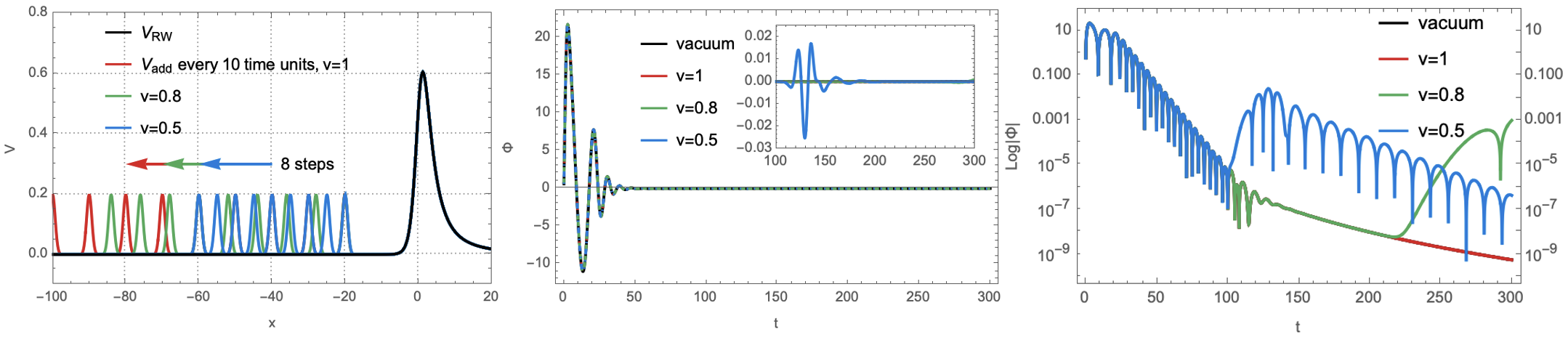}
        \caption{$a=-20$}
        \label{fig:ConstantV_n20}
    \end{subfigure}

    \begin{subfigure}[b]{0.9\textwidth}
        \centering
        \includegraphics[width=\textwidth]{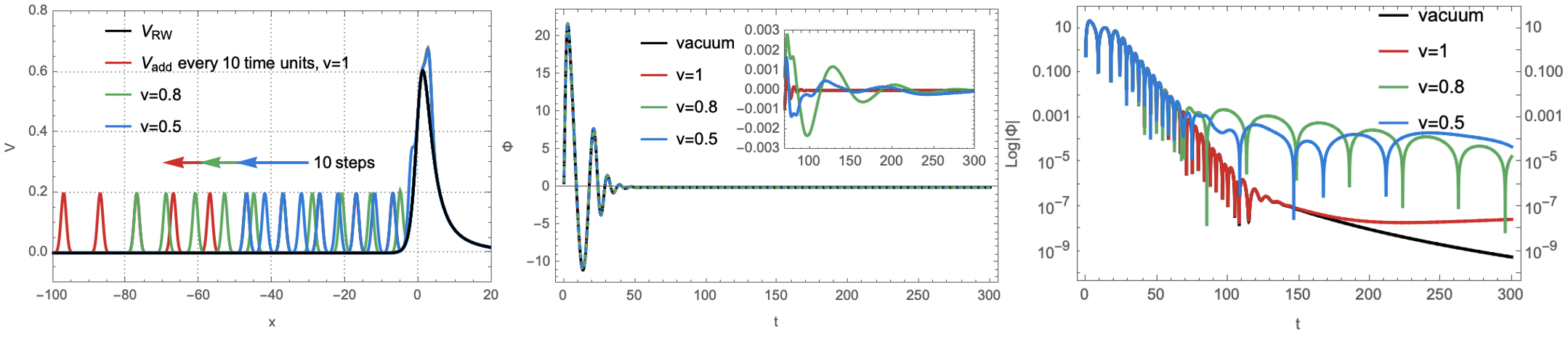}
        \caption{$a=3$}
        \label{fig:ConstantV_p3}
    \end{subfigure}

    \begin{subfigure}[b]{0.9\textwidth}
        \centering
        \includegraphics[width=\textwidth]{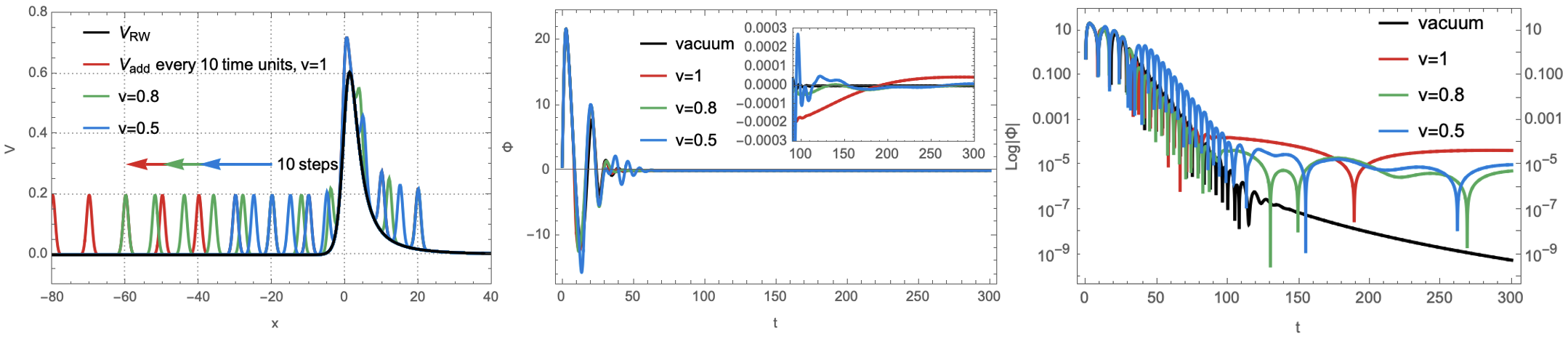}
        \caption{$a=20$}
        \label{fig:ConstantV_p20}
    \end{subfigure}

    \begin{subfigure}[b]{0.9\textwidth}
        \centering
        \includegraphics[width=\textwidth]{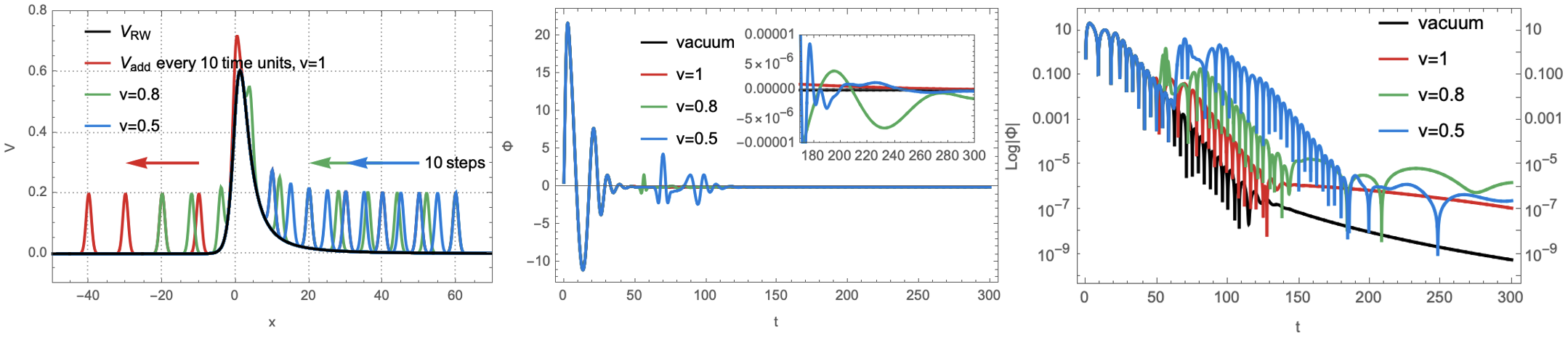}
        \caption{$a=60$}
        \label{fig:ConstantV_p60}
    \end{subfigure}
    
	\captionsetup{width=.9\textwidth}
	\caption{Waveforms of axial gravitational field perturbations with an additional bump moving at constant velocity. The blue, green, and red curves correspond to velocities $v=0.5$, $0.8$, and $1$, respectively. The black curve denotes the vacuum reference case.}
	\label{fig:moving_ConstantV}
\end{figure}

The perturbation signal, initially excited near the light ring $r=3M$, propagates both inward and outward at the speed of light. 
Whether this signal can interact with the moving bump depends on both the bump’s initial position $a_0$ and its velocity $v$. 
In Fig.~\ref{fig:moving_ConstantV}, we analyze these interactions for various initial positions ($a_0 = -60, -20, 3, 20, 60$) and constant velocities ($v = 1, 0.8, 0.5$). 
The arrows and markers in the first column illustrate the Gaussian bump trajectories at fixed time intervals.
%\red{In table.~\ref{tab:ConstantV_bump_quantitative}, we summarize all the echo details in the geodesic motion model. With the bump initially positioned left of the Regge–Wheeler potential ($v<1$), echo intervals significantly lengthen, while with the bump positioned right of the RW potential ($v<1$), echo intervals shorten and they are substantially reduced or absent.}
Table~\ref{tab:ConstantV_bump_quantitative} summarizes all echo characteristics in the geodesic motion model. When the bump is initially located to the left of the Regge–Wheeler potential ($v<1$), the echo intervals become significantly longer. Conversely, when the bump is placed to the right of the potential ($v>1$), the intervals shorten, and the echoes are greatly diminished or completely disappear.

\begin{table}[htbp]
\centering
\caption{Echoes for the constant velocity motion model}
\label{tab:ConstantV_bump_quantitative}
\begin{tabular}{cccccc}
\toprule
\multirow{2}{*}{$a$} & \multirow{2}{*}{$v$} & echo & time of appearance & amplitude ratio & \multirow{2}{*}{supplement} \\
                     &  & number & ($t/M$) & (\%) &  \\
\midrule
\multirow{3}{*}{-60} & 1 & — & — & — & — \\
                     & 0.8 & — & — & — & — \\
                     & 0.5 & 1 & 249.8 & 0.135 & — \\
\midrule
\multirow{3}{*}{-20} & 1 & — & — & — & — \\
                     & 0.8 & 1 & 221.5 & — & — \\
                     & 0.5 & 1 & 100.6 & 0.129 & frequency downshift after the echo \\
\midrule
\multirow{3}{*}{3} & 1 & — & — & — & slower decay in late-time tail \\
                   & 0.8 & — & — & — & frequency downshift during oscillation \\
                   & 0.5 & — & — & — & irregular oscillation in late-time tail \\
\midrule
\multirow{3}{*}{20} & 1 & — & — & — & slower decay in late-time tail \\
                    & 0.8 & — & — & — & irregular oscillation in late-time tail \\
                    & 0.5 & 1 & 30.6 & 11.74 & irregular oscillation in late-time tail \\
\midrule
\multirow{5}{*}{60} & 1 & 1 & 48.4 & 0.348 & slower decay in late-time tail \\
                    & \multirow{2}{*}{0.8} & \multirow{2}{*}{2} & 51.5 & 7.38 & \multirow{2}{*}{irregular oscillation in late-time tail} \\
                    &  &  & 62.6 & 1.068 &  \\
                    & \multirow{2}{*}{0.5} & \multirow{2}{*}{2} & 62.2 & 19.79 & \multirow{2}{*}{irregular oscillation in late-time tail} \\
                    &  &  & 80.4 & 11.95 &  \\
\bottomrule
\end{tabular}
\end{table}

Figs.~\ref{fig:ConstantV_n60} and \ref{fig:ConstantV_n20} show that for $v=1$ (red), the bump moves at light speed and quickly outpaces the perturbation signal, resulting in no echoes, similar to geodesic cases with asymptotically luminal motion. 
When $v < 1$, the signal eventually catches up with the bump, reflects back, and generates echoes.
This behavior is clearly seen for $v=0.5$ (blue) and $v=0.8$ (green) in $a_0=-60$ and $-20$ cases.

Interestingly, echoes induced by slower bumps also exhibit a noticeable downshift in the frequency of the decay-stage oscillations, as evidenced in Fig.~\ref{fig:ConstantV_p3}. 
In this case $a_0=3$, the waveform for $v=1$ shows the same slower power-law tail as that seen in Fig.~\ref{fig:Geodesic_p3}, resulting from the Gaussian bump's modification of the Regge–Wheeler potential profile. 
In Fig.~\ref{fig:ConstantV_p20}, $a_0=20$, a single early echo is observed for $v=0.5$, but no echo appears for $v=1$ and $0.8$, which is because the bump quickly moves away before a significant reflection occurs.
For $a_0=60$, as shown in Fig.~\ref{fig:ConstantV_p60}, all velocities produce one or more early echoes. 
And the slower the bump moves, the later these echoes appear, since the cavity between the bump and the Regge-Wheeler potential peak persists longer.
Moreover, when the bump starts to the right of the main peak, the continuous interaction during the tail phase induces distortions in the decay oscillations, highlighting the intricate dynamics between the moving barrier and the trapped perturbation signal.

In summary, the constant velocity model provides a controlled framework to systematically explore how the bump velocity and initial position influence echo production, frequency shifts, and late tail modulation. 
This complements the geodesic-inspired analysis and emphasizes the rich phenomenology arising from dynamic modifications to the effective potential.

%%%%%%%%%%%%%%%%%%%%%%%%%%%%%%%%%%%%%%%%%%%%%%%%%%%%%%%%%%%%
\section{Conclusion and Discussion}
\label{sec:concl}
%%%%%%%%%%%%%%%%%%%%%%%%%%%%%%%%%%%%%%%%%%%%%%%%%%%%%%%%%%%%

In this work, we systematically investigate the impact of an additional dynamic potential on the gravitational perturbation waveforms of black holes. 
We first construct a time-independent Gaussian bump as an effective model to represent the matter fields or exotic structures around a black hole. 
By introducing motion parameters into this bump, we explore two physically distinct scenarios: motion along geodesics and uniform motion with a constant velocity.

In the geodesic model, the bump is assumed to trace the trajectory of infalling matter moving along timelike geodesics. 
We derive the explicit expressions for the velocity profile in terms of both the radial and tortoise coordinates, and examine the dependence on initial conditions, such as initial energy and position. 
Numerical simulations demonstrate that when the bump moves with an asymptotically luminal velocity, it eventually outruns the perturbation signal, suppressing the production of echoes. 
In particular, for the cases with non-zero initial radial velocity, the bump rapidly approaches the black hole, leaving minimal interaction time for the perturbation signal. 
Conversely, when the bump starts with zero initial velocity and from a finite radius, a gradual acceleration leads to characteristic waveform modulations, including changes in the decay phase and mild amplitude differences in the late-time tails.

In contrast, the constant velocity model provides a simplified framework for controlling the bump dynamics explicitly. 
By fixing the bump velocity to subluminal values ($v < 1$), we ensure that the perturbation signal can eventually catch up with the moving bump, allowing for repeated reflections and the emergence of distinct echoes. 
Our analysis shows that the number, amplitude and timing of echoes strongly depend on both the bump velocity and its initial position relative to the Regge–Wheeler potential peak. 
Moreover, the subluminal motion introduces frequency downshifts in the late-time oscillations and leads to complex modulations in the decay tails, reflecting the intricate interplay between the moving barrier and the trapped perturbation signal.

The comparative study between the geodesic and constant velocity scenarios highlights the crucial role of the bump dynamics in shaping the gravitational waveform. 
While geodesic model is more physically motivated by astrophysical infall processes, the constant velocity model serves as a controlled toy model to systematically probe parameter dependencies. 
Both approaches illustrate how additional dynamic potential structures can significantly modify late-time gravitational signals, potentially offering new observational imprints beyond the already-known ringdown phase.

Looking ahead, our results open up intriguing possibilities for exploring nontrivial environments around black holes, such as matter clouds, dynamical hairs, or exotic compact objects with modified effective potentials~\cite{Cardoso:2019mes}. 
Further studies incorporating back-reaction effects, more realistic matter distributions, and full nonlinear dynamics are essential to assess the detectability of these signatures in gravitational wave observations. 
Additionally, it would be interesting to investigate the interplay between such dynamic matter fields and potential beyond-general-relativity effects, which might manifest as deviations from the general-relativity's quasinormal mode spectrum.

In summary, the incorporation of an additional moving potential enriches the phenomenology of black hole perturbations, offering novel mechanisms for echo production and late-time waveform modulation. 
Our findings provide a new perspective for interpreting future gravitational wave data and probing the near-horizon structure of black holes.

%%%%%%%%%%%%%%%%%%%%%%%%%%%%%%%%%%%%%%%%%%%%%%%%%%%%%%%%%%%%
\section*{Acknowledgements}
%%%%%%%%%%%%%%%%%%%%%%%%%%%%%%%%%%%%%%%%%%%%%%%%%%%%%%%%%%%%

Y.-L.\ T. and H.\ Y. contributed equally to this work.
This work was supported in part by the National Natural Science Foundation of China under Grant No.~12175108. L.C. was also supported by Yantai University under Grant No.\ WL22B224.

\appendix

\section{Physical interpretation of the bump}
\label{app:physical_interp}
%%%%%%%%%%%%%%%%%%%%%%%%%%%%%%%%%%%%%%%%%%%%%%%%%%%%%%%%%%%%

We interpret the bump by an effective potential that represents a localized matter distribution modeled here as an ideal anisotropic fluid.\footnote{The additional potential term can also be interpreted as an effective quantum correction arising from one-loop backscattering or vacuum polarization effects near the potential barrier \cite{Beltran-Palau:2022nec}, which slightly modifies the propagation of perturbations without altering the background geometry.
If the bump is interpreted as an effective quantum correction, our analysis indicates that its amplitude would be extremely small $\epsilon\lesssim 10^{-2}$, typically several orders of magnitude below the level required to generate observable echoes --- so the pronounced echo signals discussed here should instead be regarded as the phenomenological upper limit of such effects.}
The spherically symmetric spacetime metric containing such a matter field is written as
\begin{equation}
\dif s^2 = - \me^{2\alpha(r)}\dif t^2 + \me^{2\beta(r)}\dif r^2 + r^2\dif \Omega^2,
\end{equation}
where $\alpha(r)$ and $\beta(r)$ are metric shape functions, and $\me^{-2\beta(r)} = 1 - \frac{2m(r)}{r}$.
The energy–momentum tensor of the anisotropic fluid takes the form,
\begin{equation}
T^\mu_\nu = \mathrm{diag}(-\rho , p_r , p_t , p_t),
\end{equation}
where $\rho(r)$, $p_r(r)$, and $p_t(r)$ denote the energy density, radial pressure, and tangential pressure, respectively.
These quantities satisfy
\begin{equation}
m' = 4\pi r^2\rho , \qquad
\alpha' = \frac{m + 4\pi r^3p_r}{r(r-2m)} , \qquad
p_r' + (p_r + \rho)\alpha' + \frac{2}{r}(p_r - p_t) = 0 ,
\label{eqs:matter}
\end{equation}
where the prime means the derivative with respect to the radial coordinate and the third equation is the Tolman–Oppenheimer–Volkoff (TOV) equation.
Since these three equations constrain five unknown functions $\big\{\alpha(r), m(r), \rho(r), p_r(r), p_t(r)\big\}$, two additional relations are required to close the system.

The axial perturbation equation and its effective potential in this general static and spherically symmetric spacetime are given by [47]%~\cite{Chandrasekhar:1991fi}
\begin{equation}
\frac{\partial^2 \Phi}{\partial x^2} - \frac{\partial^2 \Phi}{\partial t^2} - V_{\mathrm{axial}}\Phi = 0,
\end{equation}
\begin{equation}
V_{\mathrm{axial}}(r) = \me^{2\alpha(r)}\left[\frac{\ell(\ell+1)}{r^2} - \frac{6m(r)}{r^3} + 4\pi(\rho - p_r)\right],\label{axialpoten}
\end{equation}
where the tortoise coordinate $x$ is defined through $\dif x = \me^{\beta-\alpha}\dif r$.

To analyze the perturbations, we expand the effective potential $V_{\mathrm{axial}}$ and the metric function $\alpha(r)$ to their first orders,
\begin{eqnarray}
V_{\mathrm{axial}} &=& V_0 + \delta V, \label{deltaV} \\
\alpha &=& \alpha_0 + \delta\alpha,	\label{deltaalpha}
\end{eqnarray}
where
\begin{equation}
V_0 = V_{\mathrm{RW}},
\qquad
\delta V = V_{\mathrm{add}},
\end{equation}
and
\begin{equation}
\me^{2\alpha_0} = 1-\frac{2M}{r},
\qquad
m = m_0 + \delta m,
\qquad
m_0 = M,\label{paraalphazero}
\end{equation}
and treat the matter components as first-order quantities,
\begin{equation}
\rho = \delta\rho, \qquad p_r = \delta p_r, \qquad p_t = \delta p_t .\label{threedeltas}
\end{equation}
When substituting Eqs.~(\ref{axialpoten}), (\ref{deltaalpha})-(\ref{threedeltas}) into Eq.~(\ref{deltaV}) and keeping only first-order terms, we obtain
\begin{equation}
V_{\mathrm{add}} = \delta V = 2\me^{2\alpha_0} \left[\frac{l(l+1)}{r^2} - \frac{6M}{r^3}\right] \delta\alpha - \frac{6}{r^3} \me^{2\alpha_0} \delta m + 4\pi \me^{2\alpha_0}(\rho - p_r).
\label{eq:first_order}
\end{equation}

We assume that the additional potential originates solely from the matter term,
\begin{equation}
V_{\mathrm{add}} =4\pi \me^{2\alpha_0}(\rho - p_r),\label{addpotential}
\end{equation}
implying negligible metric corrections, i.e., $\delta\alpha \simeq 0$ and $\delta m \simeq 0$.
Under this approximation, we have $\me^{2\alpha} \simeq 1-\frac{2M}{r}$ and $ m \simeq M $, which implies that the background geometry effectively reduces to the Schwarzschild spacetime and provides the two additional relations required to determine the five functions mentioned above.
For simplicity, we introduce an equation of state of the form, 
\begin{equation}
p_r(r) = \chi\rho(r),\label{equstate}
\end{equation}
where $\chi$ is a constant (typically $\chi < 1$) or a function of $r$.
From Eqs.~(\ref{addpotential}), (\ref{equstate}), and (\ref{paraalphazero}), we derive the energy density of the fluid,
\begin{equation}
\rho(r) = \frac{r}{4\pi(1-\chi)(r-2M)}V_{\mathrm{add}}(r),
\end{equation}
and rewrite the radial pressure as follows,
\begin{equation}
p_r(r) = \chi\rho(r) = \frac{\chi\, r}{4\pi(1-\chi)(r-2M)}V_{\mathrm{add}}(r).
\end{equation}
Using Eq.~(\ref{eqs:matter}), we further obtain the tangential pressure,
\begin{eqnarray}
p_t(r) &=& \frac{r}{2}(\chi\rho)' + \left[\frac{(1+\chi)M}{2(r-2M)} + \chi\right]\rho \nonumber \\
&=& \frac{\chi r^2}{8\pi(1-\chi)(r-2M)}V_{\mathrm{add}}' + \frac{((2r - 5M)\chi + r(r - 2M)\chi' + M)r}{8\pi(1-\chi)(r-2M)^2}V_{\mathrm{add}}.
\end{eqnarray}

\begin{figure}[!ht]
\centering
\includegraphics[width=\textwidth]{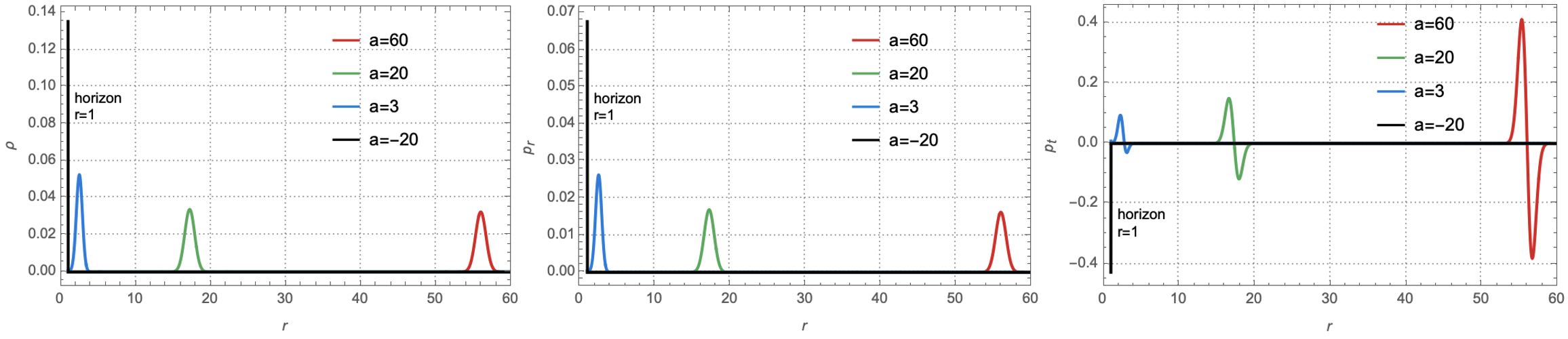}
\captionsetup{width=.9\textwidth}
\caption{Profiles of the matter functions $\rho(r)$, $p_r(r)$, and $p_t(r)$
as functions of the radial coordinate. The red, green, blue, and black curves correspond to bump positions $a = 60, 20, 3, -20$, respectively. A severe divergence occurs for the value of $a = -60$ at $r \rightarrow 1$, this situation is consequently omitted. Here $\chi=0.5$ and $2M = 1$ are set.}
\label{fig:matter_functions}
\end{figure}

As shown in Fig.~\ref{fig:matter_functions}, the energy density $\rho(r)$ and radial pressure $p_r(r)$ remain positive throughout the region of parameter $a$ we choose.
Their magnitudes remain small far from the event horizon at $r=2M$, but rise sharply and diverge near the horizon due to the $(r-2M)^{-1}$ factor.
The tangential pressure $p_t(r)$ can take both positive and negative values, but the positive regions dominate in the spatial extent.
Its magnitude decreases near the horizon; however, the analytical expressions still indicate a weak divergence close to $r=2M$, possibly reflecting matter accumulation near the black hole.

Consequently, the energy conditions for $\chi=0.5$ can be examined from the numerical results of Fig.~\ref{fig:matter_functions} as follows:
\begin{itemize}
\item \textbf{Null Energy Condition (NEC):}
$\rho(r) + p_r(r) = (1 + \chi)\rho(r) \ge 0$.
Although $\rho(r) + p_t(r)$ may be locally negative, its integral remains positive; thus, the NEC is satisfied~\cite{Curiel:2014zba,Kontou:2020bta}.
\item \textbf{Weak Energy Condition (WEC):}
$\rho(r) \ge 0$; the WEC is satisfied.
\item \textbf{Strong Energy Condition (SEC):}
$\rho(r) + p_r(r) + 2p_t(r)$ may exhibit local violations but yields a positive integrated value, indicating that the SEC holds.
\item \textbf{Dominant Energy Condition (DEC):}
$\rho(r) - p_r(r) = (1 - \chi)\rho(r) \ge 0$.
While $\rho(r) - p_t(r)$ may become negative in localized regions, its integrated value is positive; hence, the DEC is fulfilled.
\end{itemize}

For all the above calculations we adopt $\chi = 0.5$.
In general, both the pointwise and integral forms of the energy conditions remain valid for $\chi \in [-1,1]$.

Finally, we make a comment on
the mass correction $\delta m(r)$,
\begin{eqnarray}
\delta m(r) &=& 4\pi \int^r_{2M} \rho(r)r^2\dif r \nonumber \\
&=& \int^r_{2M} \frac{r^3}{(1-\chi)(r-2M)}V_{\mathrm{add}}(r)\dif r,
\end{eqnarray}
which cannot be evaluated analytically. However, the integral actually converges due to the rapid decay of $V_{\mathrm{add}}$ at the lower limit of integration. Therefore, the flexibility in choosing $\chi$ under the energy-condition constraints ensures the self-consistency of the assumption of $\delta m \simeq 0$.

\bibliographystyle{utphys}
\bibliography{references}

\end{document}